\documentclass[twocolumn,showpacs,amsmath,amssymb,prl]{revtex4}

\usepackage{graphicx}
\newcommand{\bdm}{\begin{displaymath}}
\newcommand{\edm}{\end{displaymath}}
\newcommand{\be}{\begin{equation}}
\newcommand{\ee}{\end{equation}}
\newcommand{\ba}{\begin{eqnarray}}
\newcommand{\ea}{\end{eqnarray}}
\newcommand{\eref}[1]{(\ref{#1})}

\begin{document}
\title{Feedback effects on the current correlations in
Y-shaped conductors}
\author{Shin-Tza Wu$^1$ and Sungkit Yip$^2$}
\affiliation{
$^1$Department of Physics, National Chung-Cheng University,
Chiayi 621, Taiwan\\
$^2$Institute of Physics, Academia Sinica, Nankang, Taipei 115,
Taiwan}
\date{\today}

\begin{abstract}
We study current fluctuations in a Y-shaped conductor connected
to external leads with finite impedances.
We show that, due to voltage
fluctuations in the circuit,
the moments of the transferred charges
 cannot be obtained from simple rescaling of
the bare values already in the second moments.
The cross-correlation between the output terminals can
change from negative to positive under certain parameter regimes.
\end{abstract}
\maketitle

Current fluctuations in mesoscopic systems are of
intense interest recently.  Besides
the fact that they inevitable exist and become
more important in electric circuits when these are
miniaturized, it is now well-appreciated
that they contain interesting and fundamental physics
(see, e.g., \cite{BB,BS,Nazarov_rev}).
For example, Fermi statistics of the electrons are responsible
for the reduction of the shot noise as compared with
a corresponding system of classical or Bosonic particles.
Furthermore, the same Fermi statistics of the
charge carriers have important implications
 in multi-terminal setups. The quantity
of interest in this case is the cross-correlation between different
terminals \cite{BB,Buttiker90,Martin92,MRS}.
Experiment of this type is the
solid-state analogue of the Hanbury Brown-Twiss experiment in
quantum optics \cite{QO}.
It has been shown that generally Fermi statistics
implies that the cross-correlations between the currents
in two different output arms are negative
\cite{Buttiker90,Martin92,MRS}.
This theoretical result assumes non-interacting electrons
and is supported by experiments available
so far \cite{HBT}.

The sign of this cross-correlation has triggered
many investigations into the question as to under what
circumstances it would be reversed.  Several
mechanisms have been proposed.
A few invoke possible electronic ground states
that are not normal-metallic (Landau-Fermi liquid) states,
such as superconducting (e.g.~\cite{SC}), quantum Hall,
or Luttinger liquids states (e.g.~\cite{QH}).
Still another mechanism has been proposed relying
on finite frequency and capacitive couplings \cite{MB00},
and one in ferromagnetic systems based on ``bunching" of transferred
electrons due to spin blockade \cite{CBB}.

Here we show that a feedback mechanism in the presence of
external impedances can also lead to this sign change.
 We consider the system
as shown in Fig.~\ref{circuit}.
The ``sample" $A$ is a Y-shaped conductor in which electrons
propagate coherently.  It can consist of
tunnel barriers or diffusive conductors, and is
connected to external resistors
$Z_a$, $Z_b$ and $Z_c$.  Our considerations are also
of  relevance for practical measurements
({\it c.f.}~\cite{RSP}).
 In measurements of cross-correlations, one injects an
incident beam of charge carriers (here from reservoir $a$)
and then splits the beam into
two parts using a ``beam splitter", such as the
 Y-shaped conductor here.  One would like to measure
the current correlation between two output terminals,
here $b$ and $c$.  In most current measurements however,
one needs to couple the sample to external
measuring circuits.  For example, here we are
considering the case where the current measurements
are actually made by voltage measurements across
the impedances $Z_b$ and $Z_c$.
 If the external measuring circuit can be idealized as
having zero
impedance, then the voltage across the sample would be
non-fluctuating and the current fluctuations are entirely due
to intrinsic properties of the sample and the
carriers. With finite external impedances, the voltage across the sample then
becomes fluctuating and the current correlations
will be modified.

Previously, feedback due to the presence of external
impedances has been considered for two-terminal
conductors \cite{BB}, and more recently,
in the context of third moment of
the shot noise \cite{RSP,KNB,Kind,BKN}.
The results based on the Langevin formalism \cite{BB}
concluded that the second moments of current fluctuations can be obtained
from the corresponding zero-impedance
(intrinsic or ``{\em bare}") values by a simple scaling.
However, it was shown, using both a Keldysh technique
\cite{KNB,Kind} and the Langevin
formalism \cite{Kind,BKN}, that this rescaling breaks down at the third moment.
In this work we show that, for our three-terminal setup,
 even the second moment cannot be obtained
from a rescaling of the corresponding bare value.
 For instance, the cross-correlation
acquires contributions from auto-correlators. Since the bare
auto-correlators are always positive, it is then possible to have
{\em positive} cross-correlations in  appropriate parameter regimes.
 The effect of external impedances
on current fluctuations of multi-terminal circuits has also
been considered by B\"uttiker and his collaborators \cite{BB} using
the Langevin formalism.
However, they considered a multiprobe measurement
of a two-terminal conductor and thus not directly our geometry here.

We have performed the calculations using both the
Langevin and Keldysh formalisms \cite{WY}.  The results are
identical.
To illustrate the physics more clearly, let us first
consider a simplified case where only $Z_b \ne 0$
 using the Langevin formalism. Let us first introduce some
short-hand notations.  We denote
 the conductances of the three arms of our sample
as $G_a$, $G_b$, $G_c$.  We shall define
$G \equiv G_a + G_b + G_c$ and also the dimensionless
parameters $\eta_a \equiv G_a/G$ etc (thus
$\eta_a + \eta_b + \eta_c = 1$).
In the present situation, the potentials $\phi_1$ and $\phi_3$
at points $1$ and $3$ are given by the
external potentials $V_a$ and $V_c$  respectively
and do not fluctuate.  However,
the quantity $\phi_2$ is a fluctuating quantity.
 The current in an arm b, say,
is a linear combination of two contributions,
one being linear in the bias potentials $\phi$'s
and another due to
the Langevin noise.  Thus we have
\be
I_b = G \eta_b \left[ \eta_a (V - \phi_2) - \eta_c \phi_2 \right]
  + \delta I_2 \, .
\label{Ib}
\ee
The first term follows easily
from circuit theory.  $\delta I_2$ is the Langevin
noise whose expectation value is zero.  We shall
specify its variance later.
Similarly for arm $c$,  we have
\be
I_c = G \eta_c \left[ \eta_a V  + \eta_b \phi_2 \right]
  + \delta I_3 \, .
\label{Ic}
\ee
The fluctuating potential $\phi_2$ is related to $I_b$ by
\be
\phi_2 = I_b Z_b \, .
\label{phi2}
\ee
(We are interested in the zero frequency limit so
the current along arm $A_b$ is equal to that
through the resistor $Z_b$.)
By taking the expectation values of (\ref{Ib})-(\ref{phi2}),
 we can obtain $\overline{I_b}$, $\overline{I_c}$
and $\overline{\phi_2}$.  In particular, we have
\be
\overline{\phi_2} = \frac{1}{\tilde z_t} Z_b G \eta_a \eta_b V
\, .
\label{aphi2}
\ee
Here
$\tilde z_t \equiv 1 + Z_b G \eta_b (\eta_a + \eta_c) $ is
a dimensionless quantity.
Subtracting the expectation values of Eqs.~(\ref{Ib})-(\ref{phi2})
 from these equations themselves, we find,
by eliminating $ \phi_2 - \overline{\phi_2}$ in favor
of $I_b - \overline{I_b}$,
\begin{eqnarray}
\Delta I_b &\equiv& I_b - \overline{I_b}
= \frac{1}{\tilde z_t} \delta I_2 \, ,
\label{DIb}
\\
\Delta I_c &\equiv& I_c - \overline{I_c} =
   \frac{1}{\tilde z_t} Z_b G \eta_b \eta_c \delta I_2
      + \delta I_3 \, .
\label{DIc}
\end{eqnarray}
From these, we can readily obtain the fluctuations
$\langle\Delta I_b \Delta I_c\rangle$ etc.
(We leave out the frequency variables for simplicity here.
See below for more accurate notations). In particular,
\be
\langle\Delta I_b \Delta I_c\rangle  =
\frac{1}{\tilde z_t^2} Z_b G \eta_b \eta_c
\langle\delta I_2 \delta I_2\rangle
  + \frac{1}{\tilde z_t}  \langle\delta I_2 \delta I_3\rangle\, .
\label{cross1}
\ee
This shows immediately that the cross-correlation has several
contributions.  Besides one which is a rescaling
of the ``bare" correlator $\langle\delta I_2 \delta I_3\rangle$,
there is another contribution being proportional
to $\langle\delta I_2 \delta I_2\rangle$. The origin of this latter term
is obvious also from the above derivation, that is,
the sample is ``driven" by the potential
$\phi_2$ which is itself fluctuating.
To complete the calculation we need the expressions
for $\langle\delta I_2 \delta I_2\rangle$ and
$\langle\delta I_2 \delta I_3\rangle$.
For this, we have to notice that the sample is now
biased at voltages $V$ at point $1$, $\phi_2$ at point $2$, and
$0$ at point $3$.
Let us define the {\it bare} (superscript $^{(0)}$) correlators
$C^{(0)}_{bc}$ by the expression
$\langle\delta I_2(\omega) \delta I_3 (\omega')\rangle
  =  2 \pi \delta (\omega + \omega') C^{(0)}_{bc} $
etc. In the shot noise (temperature $T \to 0$) regime, we expect these
correlators to be linear combinations of contributions
that are proportional to the average potential differences,
{\it i.e.},
\begin{eqnarray}
 C^{(0)}_{bb} / e  =
 s^{(b)}_{bb} ( V - \overline{ \phi_2 }) + s^{(c)}_{bb} ( V - 0)
\, , \label{dI2} \\
 C^{(0)}_{bc} / e =
  s^{(b)}_{bc} ( V - \overline{\phi_2 }) + s^{(c)}_{bc} ( V - 0)
\, . \label{dI23}
\end{eqnarray}
The values of the coefficients $s^{(b)}_{bb}$ etc will
be given below.
Here the superscripts are denoted according to the potentials
relative to $a$ and the subscripts, the currents.
Writing $\langle \Delta I_b(\omega)\Delta I_c(\omega')\rangle = 2 \pi
\delta(\omega+\omega') C_{bc}$,  we can now obtain the ``renormalized"
correlator $C_{bc}$ (which is also proportional to the correlators for
transferred charges) from Eq.~(\ref{cross1}),
 using (\ref{aphi2}), (\ref{dI2}) and (\ref{dI23}):

\begin{widetext}
\be
C_{bc} =  \left\{
\frac{ (Z_b G \eta_b \eta_c) ( 1 + Z_b G \eta_b \eta_c )}{\tilde z_t^3}
  s^{(b)}_{bb}
 +  \frac{ ( 1 + Z_b G \eta_b \eta_c )}{\tilde z_t^2}
  s^{(b)}_{bc}    +  \frac{ (Z_b G \eta_b \eta_c)}{\tilde z_t^2}
  s^{(c)}_{bb} +  \frac{1}{\tilde z_t} s^{(c)}_{bc}
\right\} eV \, .
\label{Cbc1}
\ee

The modifications needed for our general case are
straight-forward in principle.
 We simply state our final results:
\begin{eqnarray}
C_{bb} &=& \frac{eV}{z_t^3}
\left\{
(P+S)
\left[P^2 s_{bb}^{(b)}+Q^2 s_{cc}^{(b)}+2PQ s_{bc}^{(b)}\right]
+ (Q+R)
\left[P^2 s_{bb}^{(c)}+Q^2 s_{cc}^{(c)}+2PQ s_{bc}^{(c)}\right]
\right\} \, ,
\label{Cbb} \\
C_{cc} &=& \frac{eV}{z_t^3}
\left\{
(P+S)
\left[S^2 s_{bb}^{(b)}+R^2 s_{cc}^{(b)}+2SR s_{bc}^{(b)}\right]
+ (Q+R)
\left[S^2 s_{bb}^{(c)}+R^2 s_{cc}^{(c)}+2SR s_{bc}^{(c)}\right]
\right\} \, ,
\label{Ccc} \\
C_{bc}  &=& \frac{eV}{z_t^3}
\left\{
(P+S)
\left[PS s_{bb}^{(b)}+QR s_{cc}^{(b)}+(PR+QS) s_{bc}^{(b)}\right]
+ (Q+R)
\left[PS s_{bb}^{(c)}+QR s_{cc}^{(c)}+(PR+QS) s_{bc}^{(c)}\right]
\right\} \, .
\label{Cbc}
\end{eqnarray}
\end{widetext}
In these equations
\begin{eqnarray}
z_t &\equiv&
 \eta_a ( 1 + G Z_b \eta_b)( 1 + G Z_c \eta_c)
\nonumber \\
&+& \eta_b ( 1 + G Z_c \eta_c)( 1 + G Z_a \eta_a)
\nonumber \\
&+& \eta_c ( 1 + G Z_a \eta_a)( 1 + G Z_b \eta_b)
\end{eqnarray}
is a dimensionless number, and the symbols $P,Q,R,S$ stand for
\begin{eqnarray}
P &=& 1+ Z_a G \eta_a \eta_c + Z_c G \eta_c (\eta_a+\eta_b)
 \, ,
\nonumber \\
Q &=& -Z_a  G \eta_a \eta_b
+ Z_c G \eta_b \eta_c \, ,
\nonumber \\
R &=& 1+Z_a G \eta_a \eta_b
+Z_b G \eta_b (\eta_a + \eta_c) \, ,
\nonumber \\
S &=& -Z_a G \eta_a \eta_c
+ Z_b G \eta_b \eta_c \, .   \label{PQRS}
\end{eqnarray}
The coefficients $s_{bb}^{(b)}$ etc in
Eqs. (\ref{Cbb})-(\ref{Cbc}) are the same as
those entering Eqs.~(\ref{dI2}), (\ref{dI23}).
 Generalization of the intrinsic (``bare")
correlation between arms $\alpha$ and $\beta$ in the shot noise regime is
thus ({\em c.f.}~Eqs.~\eref{dI2}, \eref{dI23})
\be
 C^{(0)}_{\alpha \beta} / e =
  s^{(b)}_{\alpha \beta} (\overline{\phi_1} - \overline{\phi_2})   +
s^{(c)}_{\alpha \beta} (\overline{\phi_1} - \overline{\phi_3}) \, .
\label{dIg}
\ee
These coefficients take different forms for tunnel
junctions and for diffusive wires.  For tunnel junctions, if
$\overline{\phi_1}\geq\overline{\phi_2}\geq\overline{\phi_3}$, they are
given by
\begin{eqnarray}
s_{bb}^{(b)} &=&
G \eta_b [ \eta_a (1-2\eta_a\eta_b)
- \eta_c (1-2\eta_b\eta_c)] \, ,
\nonumber\\
s_{bb}^{(c)} &=&
G \eta_b\eta_c (1-2\eta_b\eta_c) \, ,
\nonumber\\
s_{cc}^{(b)} &=&
-G \eta_b\eta_c(1-2\eta_a\eta_c-2\eta_b\eta_c) \, ,
\nonumber\\
s_{cc}^{(c)} &=&
G \eta_c [ \eta_a (1-2\eta_a\eta_c)
+ \eta_b (1-2\eta_a\eta_c-2\eta_b\eta_c) ]\, ,
\nonumber\\
s_{bc}^{(b)} &=&
G \eta_b\eta_c(1-2\eta_a^2-2\eta_a\eta_c-2\eta_b\eta_c) \, ,
\nonumber\\
s_{bc}^{(c)} &=&
- G \eta_b\eta_c(1-2\eta_a\eta_c-2\eta_b\eta_c) \, .
\end{eqnarray}
In the case $\overline{\phi_1}\geq\overline{\phi_3}\geq\overline{\phi_2}$ \cite{note1}, the
corresponding coefficients
can be obtained from the above expressions by exchanging
the indices $b$ and $c$. For example, $s_{cc}^{(b)}$
can be obtained from the above expression for $s_{bb}^{(c)}$ with
all indices of its right hand members making the exchange
$b\leftrightarrow c$.
For diffusive wires, if $\overline{\phi_1}\geq\overline{\phi_2}\geq\overline{\phi_3}$,
\begin{eqnarray}
s_{bb}^{(b)} &=&
\frac{G}{3} \eta_b (\eta_a-\eta_c) \, ,
\qquad
s_{bb}^{(c)} =
\frac{G}{3} \eta_b\eta_c (1+2\eta_a) \, ,
\nonumber \\
s_{cc}^{(b)} &=&
\frac{G}{3} \eta_b\eta_c (2\eta_a-1) \, ,
\qquad
s_{cc}^{(c)} =
\frac{G}{3} \eta_c (\eta_a+\eta_b) \, ,
\nonumber\\
s_{bc}^{(b)} &=&
\frac{G}{3} \eta_b\eta_c (1-2\eta_a) \, ,
\qquad
s_{bc}^{(c)} =
- \frac{G}{3} \eta_b\eta_c \, .
\end{eqnarray}
Again, for $\overline{\phi_1}\geq\overline{\phi_3}\geq\overline{\phi_2}$,
 one can get the coefficients by exchanging $b \leftrightarrow c$ in the formulas
above. These coefficients have been calculated using
generalization of the methods proposed by
Nazarov \cite{Nazarov99}. Some of these coefficients can also be
deduced from the literature (e.g.  \cite{MRS}, \cite{SC}, and \cite{Yip05}).

Equations \eref{Cbb}-\eref{PQRS} are the main
analytic results of this
paper. The auto-correlators $C_{bb}$ and $C_{cc}$
can be shown to be positive definite \cite{WY}.
We shall concentrate on the cross-correlation
$C_{bc}$ for the rest of the paper.

For $C_{bc}$,
we can show that \cite{WY} it is always negative
 if $Z_a \eta_a $ is larger than
$Z_b \eta_b$ and $Z_c \eta_c$. Hence, we shall focus on the
rest of the parameter space.
We show the results for two particular examples,
$Z_a = 0$, $Z_b = Z_c = 1/G$ in
Fig.~\ref{zz_3d}, and
$Z_a = 0$, $Z_b = Z_c = 10/G$ in Fig.~\ref{zz_10_3d}.
 We see that, for sufficiently large $Z_b$ and $Z_c$,
it is indeed possible to have positive $C_{bc}$.
The positive region starts near small $\eta_a$,
and grows with increasing $Z_b$ and $Z_c$.
Indeed,  for $Z_b$ and $Z_c$ both $\to \infty$, one
can show that the cross-correlation actually becomes positive
for any $\eta$'s.
(This sign change is not confined to $Z_a = 0$,
for more examples, see \cite{WY}.)

We can understand this behavior physically as follows.
(For more quantitative statements, see \cite{WY}.)
When there is a positive fluctuation of the current
through, say the arm $b$, there is a corresponding
increase in the potential at point $2$ in Fig.~\ref{circuit}.
This voltage fluctuation in turn will lead to
an extra current through the arm $c$, thus
giving a {\it positive} contribution to
the cross-correlation $C_{bc}$.  This contribution
will in particular be large for small $\eta_a$,
since most of this fluctuating current will
flow through $c$.  We have a net positive
$C_{bc}$ if these contributions overwhelm the
``bare" negative correlation contribution (see Eq.~(\ref{Cbc})).
In particular, since $s_{bc}^{(b)}+s_{bc}^{(c)} (<0)$ is proportional to
$\eta_a^2$ for tunnel junctions whereas it is proportional to $\eta_a$ for
diffusive wires for small $\eta_a$, it is therefore easier to get
positive $C_{bc}$ for tunnel junctions than for diffusive wires.

Our mechanism for sign change is distinct from that
due to bunching ({\em c.f.}~\cite{CBB}). We have calculated
also the Fano factors and found no
bunching in the injected current \cite{WY}.

In conclusion, we have shown that,
for a multi-terminal conductor connected
to external leads with finite impedances,
the moments of the transferred charges
 cannot be obtained from simple rescaling of the bare moments.
The cross-correlation between the output terminals can even
become positive under certain parameter regimes.

This
research was supported by NSC of Taiwan under grant numbers
NSC93-2112-M-194-019 and NSC93-2112-M-001-016.

\begin{figure}[!b]
\caption{\small Schematic for the circuit considered in this
paper. The arms $A_a$, $A_b$, $A_c$ of the Y-shaped conductor
$A$ are connected to external leads biased, respectively, at
voltages $V_a = V$, $V_b$, and $V_c$ ($V_b = V_c = 0$ in this paper).
The leads
are assumed to have impedances $Z_a$, $Z_b$, and $Z_c$, which
are schematized as external resistors connected to the sample
arms. The nodes $1,2,3$ between the sample arms and the resistors
are where voltage fluctuations set in.}
\label{circuit}
\end{figure}

\begin{figure}[!b]
\caption{\small Plots for the cross-correlations of
Y-shaped conductors with (a) tunnel junctions and (b) diffusive
wires in the arms. Here the external impedances
are $Z_a=0$ and $Z_b=Z_c=1/G$. $C_{bc}$ is in units of
$e G V$ and the thick lines over the surfaces mark the contour
$C_{bc}=0$. }
\label{zz_3d}
\end{figure}
\begin{figure}[!b]
\caption{\small Same as Fig.~\ref{zz_3d} except that
 impedances are
$Z_a=0$ and $Z_b=Z_c=10/G$.}
\label{zz_10_3d}
\end{figure}


\begin{thebibliography}{99}
\bibitem{note1} For $V_a>V_b>V_c$,
the intermediate potentials must satisfy
$\overline{\phi_1} \ge \overline{\phi_2}$
 and  $\overline{\phi_1} \ge \overline{\phi_3}$.
\end{thebibliography}
\end{document}